\documentclass[showpacs,preprintnumbers,amsmath,amssymb]{revtex4}

\usepackage{epsf}
\usepackage{psfrag}
\usepackage{epsfig}
\usepackage{graphics}
\usepackage{amsmath,amssymb,graphicx}
\usepackage{slashed}

\begin{document}

\newcommand{\be}{\begin{equation}}
\newcommand{\ee}{\end{equation}}
\newcommand{\bq}{\begin{eqnarray}}
\newcommand{\eq}{\end{eqnarray}}
\newcommand{\dt}{\frac{d^3k}{(2 \pi)^3}}
\newcommand{\dtp}{\frac{d^3p}{(2 \pi)^3}}
\newcommand{\kbruto}{\hbox{$k \!\!\!{\slash}$}}
\newcommand{\pbruto}{\hbox{$p \!\!\!{\slash}$}}
\newcommand{\qbruto}{\hbox{$q \!\!\!{\slash}$}}
\newcommand{\lbruto}{\hbox{$l \!\!\!{\slash}$}}
\newcommand{\bbruto}{\hbox{$b \!\!\!{\slash}$}}
\newcommand{\parbruto}{\hbox{$\partial \!\!\!{\slash}$}}
\newcommand{\Abruto}{\hbox{$A \!\!\!{\slash}$}}

\title{\textbf{QED with minimal and nonminimal couplings: on the quantum generation of Lorentz
violating terms in the pure photon sector}}

\date{\today}

\author{G. Gazzola$^{(a)}$} \email []{ggazzola@fisica.ufmg.br}
\author{H. G. Fargnoli$^{(a)}$} \email []{helvecio@fisica.ufmg.br}
\author{A. P. Ba\^eta Scarpelli$^{(b)}$} \email[]{scarpelli.apbs@dpf.gov.br}
\author{Marcos Sampaio$^{(a)}$} \email[]{msampaio@fisica.ufmg.br}
\author{M. C. Nemes$^{(a)}$}\email[]{carolina@fisica.ufmg.br}

\affiliation{(a) Universidade Federal de Minas Gerais - Departamento de F\'{\i}sica - ICEx \\
P.O. BOX 702, 30.161-970, Belo Horizonte - Brazil}
\affiliation{(b) Setor T\'ecnico-Cient\'{\i}fico - Departamento de Pol\'{\i}cia Federal \\
Rua Hugo D'Antola, 95 - Lapa - S\~ao Paulo - Brazil}

\begin{abstract}
We consider an effective model formed by usual QED (minimal coupling) with the
addition of a nonminimal Lorentz violating interaction (proportional to a fixed
4-vector $b_\mu$) which may radiatively generate both CPT even and odd terms in
the pure gauge sector.

We show that gauge invariance from usual QED, considered as a limit of the model
for $b_\mu \rightarrow 0$, plays an important role in the discussion of the
radiatively induced Lorentz violating terms at one-loop order. Moreover, despite
the nonrenormalizability of the (effective) model preventing us from readily
extending our discussion to higher orders, it is still possible to display the general
form of the breaking terms of the photon sector in the on shell limit
organized in powers of $b_\mu$ which in turn can be considered as a small
expansion parameter.
\end{abstract}


\maketitle

\section{Introduction}

Lorentz covariance and CPT symmetry are basic ingredients for the construction of local
Quantum Field Theories. Consequently, they are built in the Standard Model for the fundamental interactions.
However, Lorentz and CPT violating models have been subject of intensive investigation in
the past ten years \cite{kostelecky1}-\cite{ferreira} as possible limits of a more fundamental
theory. The Standard Model Extension (SME) \cite{kostelecky1}-\cite{coleman2} proposes a wide
range of possibilities in this context, concerning the gauge and fermion sectors.

In the gauge sector, the studies are concentrated mainly in two kinds of terms: the Chern-Simons-like
CPT odd term of Carroll-Field-Jackiw \cite{jackiw} and a quadratic in the field strength CPT even term
\cite{kostelecky2}.
For the fermion sector, the investigation of the Lorentz and CPT violating axial term has dominated the scenario,
mainly because, in this case, there was a controversy on the radiatively generated Carroll-Field-Jackiw term
\cite{CS1}-\cite{scarp1}.

Recently, the interest in the CPT even terms has increased due to the development of the aether concept
and to the study of extra dimensions. In \cite{carrol}, the authors discussed the use
of Lorentz violating tensor fields
(aether) with expectation values aligned with the extra dimensions in order to keep these extra dimensions hidden.
The Lorentz violating terms of the gauge, fermion and scalar sectors would come from the interaction between
these fields with the aether fields. In \cite{mgomes1} and \cite{mgomes2}, the quantum
generation of these aether interaction terms were studied. Particularly, in the case of
Electrodynamics \cite{mgomes2}, the minimal coupling term has been substituted by a nonminimal
Lorentz violating one. The aether interaction term is then generated when
the photon self-energy is calculated, with a coefficient which is regularization dependent
in the case of four spacetime dimensions.

In this paper we consider a modified version of Quantum Electrodynamics in four dimensions with the
coupling between the photon and the fermion composed by two terms: a nonminimal and the minimal one.
This model has been considered before in the papers \cite{Belich1}, \cite{Belich2} and \cite{Belich3}
in the context of Relativistic Quantum Mechanics. In \cite{Belich2} the model has been used in the calculation
of corrections for the Hydrogen spectrum and very stringent bounds have been set up in the magnitude
of the Lorentz violating parameter. In \cite{Belich3} the model has been used to study the magnetic
moment generation from the nonminimal coupling, since a tiny magnetic dipole moment of truly elementary
neutral particles might signal Lorentz symmetry violation.

Here we analyze the quantum corrections to the photon sector, considering the theory as an effective
model. This is only one among many other possible nonrenormalizable Lorentz violating terms. Besides the very
rich physical content explored at tree level in \cite{Belich1}, \cite{Belich2} and \cite{Belich3}, the motivation for the
present investigation of the  model is the interesting aspect which is the possible radiative
generation of both the Carroll-Field-Jackiw (CFJ) and CPT even terms. Although the quantum generation of the CFJ term
from an axial term in the fermion sector has been vastly investigated, the present possibility has not been
explored yet.

The fact that the model is gauge invariant allows for an interesting analysis
on the value of both the coefficients of the hypothetical CPT odd and even radiatively generated
terms.
Indeed, we show that the finite one-loop Lorentz violating corrections are prevented to be generated if a gauge invariant
regularization prescription is used. However, a gauge violating one followed by a symmetry restoring counterterm in the Lorentz
symmetric sector opens the possibility for such inductions.
Concerning higher order calculations, the very stringent bounds set up in \cite{Belich2} for the Lorentz violating parameter
strongly constrains the structure of the counterterms beyond the one-loop order.
For an effective model, the cutoff energy is a very important parameter and should be established on physical grounds.
Desired features of a Quantum Field Theory like causality and stability can be lost at very high energy \cite{causality1},
\cite{causality2}. We propose a condition the cutoff should satisfy in order to prevent the proliferation
of new terms beyond one-loop order.

This paper is organized as follows: the second section is dedicated to the presentation of the
model; the third section presents an analysis on the one-loop generation of Lorentz violating terms in the
pure gauge sector; in section four we carry out a calculation on gauge invariance grounds to fix the
coefficients of the one-loop quantum corrections; in section five we analyze on general grounds the predictability
of the model beyond the one-loop order using symmetry arguments. We draw our conclusions in section six.

\section{The model}

We begin by considering the action used in \cite{mgomes2} in four spacetime dimensions,
in which the minimal coupling term of QED
is substituted with the Lorentz violating nonminimal interaction term
$-e\bar \psi \varepsilon^{\mu \nu \alpha \beta}
\gamma_\beta b_\mu F_{\nu \alpha} \psi$, with $b_\mu$ being a constant vector that selects a fixed
direction in spacetime. This term is easily shown to be gauge invariant. Nevertheless,
the whole action is not, due to the lack of the minimal coupling. As the authors show,
when the one-loop correction to the photon two point function is calculated, a regularization dependent
aether term is generated. For such a model there is no way of fixing the coefficient
of this term.

Here we intend to analyze a modified version of this model \cite{Belich1}, \cite{Belich2}, \cite{Belich3},
in which gauge invariance
of the original action is restored by the presence of the minimal coupling term, along with the Lorentz
violating nonminimal one:
\be
\Sigma = \int d^4x \left\{ -\frac 14 F_{\mu \nu} F^{\mu \nu} + \bar \psi \left( i \parbruto - m
- e \Abruto - e \varepsilon ^{\mu \nu \alpha \beta}
\gamma_\beta b_\mu F_{\nu \alpha} \right) \psi \right\}.
\ee
The interaction term contains two contributions: the traditional QED vertex and the Lorentz
violating one. Besides the fact that $\Sigma$ is gauge invariant, we will see that its one-loop quantum
corrections may yield two kinds of Lorentz violating terms in the photon sector: the CPT even
aether term and the Carroll-Field-Jackiw CPT odd Chern-Simons-like term \cite{jackiw}.
We will discuss here the possibility of fixing the coefficients of these terms on gauge invariance grounds.

These contributions  would be generated by the vacuum polarization tensor, composed at one-loop order
by only one diagram.
Effectively, this amplitude can be decomposed in four parts, since from the calculational point of view
the problem is equivalent to the one in which there are two different vertices,
\be
\Pi^{\mu \nu}(p)=\Pi^{\mu \nu}_{11}(p)+ \Pi^{\mu \nu}_{12}(p)+\Pi^{\mu \nu}_{21}(p)+
\Pi^{\mu \nu}_{22}(p).
\ee
The lower indices refer to the vertices, where $1$ ($2$) refers to minimal (nonminimal) coupling.
In the equation above, the first term corresponds to the traditional QED, the two crossed terms
would be responsible for the Chern-Simons term and the last one would generate
the CPT even part. We now comment on the consequences of considering the complete
amplitude. The Lorentz violating three last terms are individually gauge invariant. So, focusing simply in
these pieces will not bring any light to the discussion, if we intend to use gauge symmetry in order to
fix arbitrary coefficients in our model. We enforce that we are considering here gauge invariance
of the action (transversality of the vacuum polarization tensor), a less restrictive condition which, in principle,
allows for a Chern-Simons-like term.

First, let us write all the contributions to the one-loop photon self-energy:
\begin{eqnarray}
&&\Pi_{11}^{\mu\nu}=-\text{tr}{\int_k{ie\gamma^{\nu}s(p+k)ie\gamma^{\mu}s(k)}}
\nonumber \\
&&=e^2\text{tr}\int_k{\gamma^\nu s(p+k)\gamma^\mu s(k)}=e^2 T^{\mu\nu},
\label{QED}\\
&&\Pi_{12}^{\mu\nu}=-\text{tr}{\int_k{\left(-2ie\varepsilon^{\alpha\beta\nu}_{\;\;\;\;\;\;\rho}\gamma^{\rho}
b_{\alpha}(-p_{\beta})\right)s(p+k)ie\gamma^{\mu}s(k)}}
\nonumber \\
&&=2e^2\varepsilon^{\alpha\beta\nu}_{\;\;\;\;\;\;\rho}b_{\alpha}p_{\beta} T^{\mu \rho},
\label{CS1}\\
&&\Pi_{21}^{\mu\nu}=-2e^2\varepsilon^{\alpha\beta\mu}_{\;\;\;\;\;\;\rho}b_{\alpha}p_{\beta} T^{\rho \nu}
\label{CS2}
\end{eqnarray}
and
\bq
\Pi_{22}^{\mu\nu}=-4e^2(\varepsilon^{\alpha\beta\nu}_{\;\;\;\;\;\;\rho}b_{\alpha}p_{\beta})
(\varepsilon^{\lambda\sigma\mu}_{\;\;\;\;\;\;\delta}b_{\lambda}p_{\sigma}) T^{\rho\delta},
\label{aether}
\eq
where
\begin{equation}
T^{\mu\nu}=\text{tr}\int_k{\gamma^\nu s(p+k)\gamma^\mu s(k)},
\label{A}
\end{equation}
with $s(k)$ the fermion propagator and $\int_k$ standing for $\int d^4k/\left(2 \pi\right)^4$. We
observe that, due to the Lorentz structure of $T^{\mu \nu}$, it follows that $\Pi_{12}^{\mu\nu}=\Pi_{21}^{\mu\nu}$

\section{The Lorentz violating radiative corrections}

The one-loop corrections to the photon self-energy are given by the equations (\ref{QED})-(\ref{aether}).
We turn our attention to the three last contributions, which are the possible sources of Lorentz violation
in the pure photon sector. We first consider the general form of the tensor $T^{\mu \nu}$ without calculating
explicitly the amplitude. Then we will focus on the part of interest in the amplitude to
determine the coefficients.

We begin by analyzing the two identical contributions coming from
equations (\ref{CS1}) and (\ref{CS2}), which could produce a CPT odd term, the
Chern-Simons-like term of Carroll-Field-Jackiw. We have
\be
\Pi^{\mu \nu}_{12}+ \Pi^{\mu \nu}_{21}=
4 e^2\varepsilon^{\alpha\beta\nu}_{\;\;\;\;\;\;\rho}b_{\alpha} p_{\beta} T^{\mu \rho},
\ee
in which
\be
T^{\mu \rho}=\left(p^\mu p^\rho -p^2 g^{\mu \rho} \right) \Pi (p^2) + \alpha m^2 g^{\mu \rho}
+ \beta p^\mu p^\rho
\label{Tmunu}
\ee
is the most general expression for $T^{\mu \rho}$. $\Pi(p^2)$ is a function which may embody divergences
and $\alpha$, $\beta$ are dimensionless constants. The Chern-Simons-like term
is given by
\be
\Pi_{CS}^{\mu \nu}= \lim_{p^2 \to 0} \left(\Pi^{\mu \nu}_{12}+ \Pi^{\mu \nu}_{21}\right).
\ee
So, we see that
\be
\Pi_{CS}^{\mu \nu}= 4 \alpha m^2e^2\varepsilon^{\alpha\beta\nu \mu}b_{\alpha} p_{\beta}.
\ee

For the CPT even term we perform a similar analysis:
\be
\Pi_{22}^{\mu\nu}=-4e^2(\varepsilon^{\alpha\beta\nu}_{\;\;\;\;\;\;\rho}b_{\alpha}p_{\beta})
(\varepsilon^{\lambda\sigma\mu}_{\;\;\;\;\;\;\delta}b_{\lambda}p_{\sigma}) T^{\rho\delta}.
\ee
We need now only the parts of $T^{\rho \delta}$ with $p^2=0$. So
\bq
&&\Pi^{\mu \nu}_{even}=-4 \alpha e^2m^2(\varepsilon^{\alpha\beta\nu}_{\;\;\;\;\;\;\rho}b_{\alpha}p_{\beta})
(\varepsilon^{\lambda \sigma \mu \rho}b_{\lambda}p_{\sigma}) \nonumber \\
&&=-4 \alpha e^2m^2 g^{\nu \kappa}b^\alpha p^\beta b_\lambda p_\sigma \varepsilon_{\alpha \beta \kappa \rho}
\varepsilon^{\lambda \sigma \mu \rho}.
\eq
Using the properties of the Levi-C\`ivita tensor, we perform the contractions to obtain
\bq
&&\Pi^{\mu \nu}_{even}=-4 \alpha e^2m^2 \left\{ \left(p^\mu p^\nu - p^2 g^{\mu \nu}\right) b^2 \right. \nonumber \\
&& \left. +(p \cdot b)^2 g^{\mu \nu} -(p \cdot b) \left(p^\mu b^\nu + p^\nu b^\mu\right) + p^2 b^\mu b^\nu \right\}.
\eq
The first two terms together have the Maxwell form and are not of interest. The last four contributions
give the following correction to the Lagrangian density:
\be
{\cal L}_{aether}= 2 \alpha e^2m^2 \left( b^\mu F_{\mu \nu} \right)^2,
\ee
which is the aether-like term analyzed in \cite{carrol} and generated in \cite{mgomes1} and \cite{mgomes2},
$b^\mu$ playing the role of an aether field. This term can be mapped in the classical CPT even term
proposed in \cite{kostelecky2},
\be
{\cal L}_{even}=-\frac 14 \kappa_{\mu \nu \alpha \beta} F^{\mu \nu} F^{\alpha \beta},
\ee
as long as we establish the relation
\be
\kappa_{\mu \nu \alpha \beta}= -2 \alpha e^2m^2 \left(
g_{\mu \alpha} b_\nu b_\beta - g_{\nu \alpha} b_\mu b_\beta
+ g_{\nu \beta} b_\mu b_\alpha - g_{\mu \beta} b_\nu b_\alpha
 \right).
\ee

In the following, we carry out an attempt to evaluate the $\alpha$ constant. As discussed above,
we can write $\alpha m^2 g^{\mu \nu}=T^{\mu \nu} (p=0)$. So, we have
\bq
&&\alpha m^2 g^{\mu \nu}=\int_k^\Lambda \frac{\mbox{tr}\left[ \gamma^\mu (\kbruto+m)\gamma^\nu(\kbruto+m)\right]}
{(k^2-m^2)^2} \nonumber \\
&&=4 \int_k^\Lambda \frac{2k^\mu k^\nu -(k^2-m^2)g^{\mu \nu}}{(k^2-m^2)^2} \nonumber \\
&&=4 \int_k^\Lambda \left\{ 2\frac{k^\mu k^\nu}{(k^2-m^2)^2} - \frac{g^{\mu \nu}}{(k^2-m^2)} \right\} \nonumber \\
&&=-4 \int_k^\Lambda \frac{\partial}{\partial k_\mu} \left\{ \frac{k^\nu}{(k^2-m^2)} \right\},
\eq
where the index $\Lambda$ is to indicate that the integrals are divergent and need some regularization.

The calculation above reveals that $\alpha$ is originated from a surface term coming from a difference
between two quadratically divergent integrals. Note that, as observed in \cite{mgomes1}, this coefficient
is totally regularization dependent. In the next section, we perform a more general discussion in order
to determine this constant on symmetry grounds \cite{jackiw2}.

\section{The one-loop transversal photon self-energy}

A question that naturally arises involves a possible violation of some Ward-Takahashi identity when
radiative corrections
are taken into account. In other words, is there an anomaly in the model? We show in this section that,
since conventional QED is gauge invariant, there is no room for a non transversal vacuum polarization tensor in
the present model.

Let us begin by the contraction of the external momentum with the whole amplitude. We have
\bq
&&p_\mu \Pi^{\mu \nu}= e^2p_\mu T^{\mu \nu} +
2e^2\varepsilon^{\alpha\beta\nu}_{\;\;\;\;\;\;\rho}b_{\alpha}p_\mu p_{\beta} T^{\mu \rho} \nonumber \\
&&-2e^2\varepsilon^{\alpha\beta\mu}_{\;\;\;\;\;\;\rho}b_{\alpha}p_\mu p_{\beta} T^{\rho \nu}
-4e^2(\varepsilon^{\alpha\beta\nu}_{\;\;\;\;\;\;\rho}b_{\alpha}p_{\beta})
(\varepsilon^{\lambda\sigma\mu}_{\;\;\;\;\;\;\delta}b_{\lambda}p_\mu p_{\sigma}) T^{\rho\delta}
\label{transv}
\eq
and it is easy to see that the two last terms are identically null, since the antisymmetric Levi-C\`ivita
tensor is contracted with the symmetric product $p_\mu p_\beta$. The second term is also identically
null for the same reason, since the Lorentz structure of $T^{\mu \rho}$ is the one presented in
eq. (\ref{Tmunu}). We are left with the first contribution, which is nothing but the conventional QED term. This
shows that in the case of the vacuum polarization tensor, a violation of gauge symmetry in the modified
model is only possible if the same violation occurs in the traditional QED.

We conclude this section by obtaining some conditions that regularization schemes must
satisfy in order to respect this transversality. As we will see, this is closely connected to the
generation or not of the Lorentz violating terms discussed in the last section. Let us analyze the
first term in eq. (\ref{transv}), namely
\bq
&&p_\mu T^{\mu \nu}=
\int_k{\text{tr}\left\{\gamma^\nu s(k)\;\slashed{p}\;s(k+p)\right\}} \nonumber \\
&&=-\int_k{\text{tr}\left\{\gamma^\nu\frac{1}{\slashed{k}-m}\;\slashed{p}\;\frac{1}{\slashed{k}+ \pbruto-m}\right\}}.
\eq
Using the identity $\pbruto=(\slashed{k}+\slashed{p}-m)-(\slashed{k}-m)$, we have
\bq
&&p_\mu T^{\mu \nu}=-\int_k^\Lambda{\text{tr}\left\{\gamma^\nu\frac{1}{\slashed{k}-m}\right\}}
+\int_k^\Lambda{\text{tr}
\left\{\gamma^\nu\frac{1}{\slashed{k}+\slashed{p}-m}\right\}} \nonumber \\
&&=-4 \int_k^\Lambda \frac{k^\nu}{(k^2-m^2)} + 4 \int_k^\Lambda \frac{(k+p)^\nu}{\left[(k+p)^2-m^2\right]}.
\eq
In the expression above, the first term is null and the second one differs from it only by a shift
in the momentum of integration. Since the second term is linearly divergent it differs
from the first one only by a surface term. The index $\Lambda$ is to indicate that, since the integrals
are divergent, they must be regularized. We will not use any particular regularization here. Rather, we will
maintain it implicit \cite{papersIR} as an artifact for a more general discussion. For the identification
of the surface terms, we use recursively, in the second term above, the identity
\be
\frac{1}{(k+p)^2-m^2}=\frac{1}{k^2-m^2}-\frac{p^2+2p \cdot k}{(k^2-m^2)\left[(k+p)^2-m^2\right]}
\ee
to obtain, in a simple calculation,
\be
p_\mu T^{\mu \nu}=-4p^\nu\left\{ \alpha_1+p^2(\alpha_3-2\alpha_2) \right\},
\ee
where the $\alpha_i$'s are given by
\be
\alpha_1 g_{\mu \nu} \equiv \int^{\Lambda}_k
\frac{g_{\mu\nu}}{k^2-m^2}-
2\int^{\Lambda}_k
\frac{k_{\mu}k_{\nu}}{(k^2-m^2)^2}=\int_k ^\Lambda \frac{\partial}{\partial k^\mu}
\left( \frac{k_ \nu}{(k^2-m^2)} \right),
\ee
\be
\alpha_2 g_{\mu \nu} \equiv \int^{\Lambda}_k
\frac{g_{\mu\nu}}{(k^2-m^2)^2}-
4\int^{\Lambda}_k
\frac{k_{\mu}k_{\nu}}{(k^2-m^2)^3}=\int_k ^\Lambda \frac{\partial}{\partial k^\mu}
\left( \frac{k_ \nu}{(k^2-m^2)^2} \right)
\label{CR1}
\ee
and
\be
\alpha_3 g_{\{\mu \nu}g_{\alpha \beta\}}   \equiv
g_{\{\mu \nu}g_{\alpha \beta\}}
\int^{\Lambda}_k
\frac{1}{(k^2-m^2)^2}
-24\int^{\Lambda}_k
\frac{k_{\mu}k_{\nu}k_{\alpha}k_{\beta}}{(k^2-m^2)^4},
\label{CR2}
\ee
which means
\be
g_{\{\mu \nu}g_{\alpha \beta\}}(\alpha_3-\alpha_2)=\int_k^\Lambda \frac{\partial}{\partial k^\beta}
\left[ \frac{4k_\mu k_\nu k_\alpha}{(k^2-m^2)^3} \right].
\ee

We see that in order to obtain a transversal vacuum polarization tensor there are two possibilities:
the first one fixes all the surface terms to zero ($\alpha_i=0$) and the second one fixes
$\alpha_1=0$ and $\alpha_3=2\alpha_2$. It is important to note that in the two options, $\alpha_1=0$
is a necessary condition for the preservation of gauge symmetry. Finally, we recognize here the constant
$\alpha$ of the last section, as given by the relation
\be
\alpha m^2= -4 \alpha_1.
\ee

Now it is important to carry out a discussion on this result. The values of the $\alpha_i$ constants are determined by the regularization procedure used to calculate them. So, they are regularization dependent. If a regularization technique
which preserves gauge invariance of the traditional QED is used, $\alpha_1$ will be set to zero. However, it is always possible to choose a
gauge noninvariant procedure and then restore the gauge symmetry by means of a non-symmetric counterterm. In this case, since
the Lorentz violating and Lorentz preserving sectors are independent of each other, the Lorentz breaking terms would survive.
This would be equivalent to use different
regularization procedures in the different sectors (Lorentz invariant and Lorentz violating). Nevertheless, we think the natural framework
is choosing a unique regularization for the calculation of the integrals of the same amplitude. In this case, the calculation with a transversality
preserving one would not generate, at one-loop order, these two Lorentz violating terms (CPT-odd and CPT-even).

We enforce that this is not a too strong condition for the gauge invariance which,
as a consequence, forbid the generation of the Chern-Simons term a priori. This would be the case of imposing gauge invariance of the Lagrange density, which is stronger than the gauge invariance of the action. This last one is accomplished simply
with the transversality of the photon self-energy, which in turn, allows for a Chern-Simons term. We emphasize that we have used the weaker condition, the gauge invariance of the action. The transversality of the vacuum polarization tensor fixes $\alpha_1=0$. In order to illustrate that this
is not a restrictive condition in what concerns the induction of the Chern-Simons term, we present the result of the papers \cite{scarp1} and \cite{CS8}, in which the induction of this Lorentz violating term from the axial one in the fermion sector
($b_\mu \bar\psi \gamma^\mu \gamma^5 \psi$) has been studied in the same grounds used here.
In these papers, the following result for the one-loop photon self-energy has been obtained, up to the first order in the $b_\mu$ constant
(for simplicity, in the massless case):
\bq
&&\Pi^{\mu \nu}=\Pi(p^2)(p^\mu p^\nu-p^2 g^{\mu \nu}) -4 \alpha_1 g^{\mu \nu}\nonumber \\
&&-\frac 43 \left\{ \alpha_2(p^\mu p^\nu-p^2 g^{\mu \nu}) + (2p^\mu p^\nu +p^2 g^{\mu \nu})(\alpha_3-2 \alpha_2)\right\} \nonumber \\
&&+4 i \alpha_2 p_\alpha b_\beta \epsilon^{\mu \nu \alpha \beta},
\eq
in which the $\alpha_i$'s are the same defined in the present paper.
Note that in this case, the transversality condition requires $\alpha_1=0$ and $\alpha_3=2 \alpha_2$, just like in the present study.
The difference is that in the equation above the coefficient of the Chern-Simons term is proportional to $\alpha_2$,
which is not required to be zero.
So, the procedure we adopt is not restrictive concerning the generation of the Chern-Simons term.
This result is simply a particularity of this model.

\section{Discussion on higher order contributions}

When higher order calculations are performed and internal photon lines are considered, the
degree of divergence of the contributions will increase by one unity for each nonminimal internal vertex.
We discuss in this section the capacity of prediction of the model in these cases. We are interested here
in the corrections to the photon sector. For multiloop
calculations, we can identify two kinds of contributions, distinguished by the appearance or not
of nonminimal internal vertices.

We start to examine the latter. For a n-loop calculation of the vacuum polarization tensor,
for each topology, we have to consider all combinations of the two kinds of vertices. In the case where
nonminimal vertices appear only in external points in the diagram,
we can separate them in four groups, just as in the one-loop calculation: the first, with
only minimal couplings; the second, where the nonminimal coupling occurs only in the first vertex; the third, in
which it occurs in the second external vertex; and the fourth, where the two external vertices are nonminimal.

Let us suppose that we have in the n-th order the following expression for the photon self-energy of pure QED:
\be
T_{\mu \nu}^{(n)}=T_{\mu \nu}^{(n)1}+ T_{\mu \nu}^{(n)2}+ \cdots + T_{\mu \nu}^{(n)k},
\ee
where each of the $T_{\mu \nu}^{(n)i}$ corresponds to a different graph. Let us take the second group
defined above as an example, in which all the graphs have the first vertex given by
$ie \varepsilon^{\mu \rho \sigma \lambda} b_\rho p_\sigma \gamma_\lambda$. It is easy to see
that the contribution of the second group will be given by
\be
\Pi_{\mu \nu}^{(n)21}= \varepsilon_{\mu \rho \sigma}^{\;\;\;\;\;\;\lambda} b^\rho p^\sigma T_{\lambda \nu}^{(n)}.
\ee
Among the four groups we are discussing, the second and the third will contribute to the Chern-Simons-like term.
Just as in the one-loop analysis,
\be
T_{\lambda \nu}^{(n)}=\left(p_\lambda p_\nu -p^2 g_{\lambda \nu} \right) \Pi^{(n)} (p^2)
+ \alpha^{(n)} m^2 g_{\lambda \nu} + \beta^{(n)} p_\lambda p_\nu
\label{Tmunu-n}
\ee
is the most general expression for the pure QED photon self-energy. The CPT odd term is
obtained in the limit $p^2 \to 0$ and, thus, its coefficient is proportional to $\alpha^{(n)}$. But
the total vacuum polarization
at n-loop order is given by
\be
\Pi_{\mu \nu}^{(n)}= T_{\mu \nu}^{(n)} + \Pi_{\mu \nu}^{(n)21}+ \Pi_{\mu \nu}^{(n)12} + \Pi_{\mu \nu}^{(n)22}
+ I_{\mu \nu}^{(n)},
\ee
where $I_{\mu \nu}^{(n)}$ includes all the graphs with internal nonminimal couplings.
It is easy to see that in order
to have $p^\mu T_{\mu \nu}^{(n)}=0$, $\alpha^{(n)}=0$ is an unavoidable condition, like in the one-loop case.
Thus, the graphs in which the nonminimal couplings occur only with external photons do not contribute to the
radiative generation of the Lorentz violating CPT odd term if a gauge invariant regularization
procedure is used. The same argument is easily extended for the
CPT even term. These terms can be radiatively generated, therefore, only with the contribution of the
nonminimal coupling to the internal photons.

Now the next step is to consider the contributions included in $I_{\mu \nu}^{(n)}$ with the occurrence of
nonminimal couplings to internal photons. Again, we consider the generation of the CPT odd term, since
a similar analysis can be performed for the CPT even one. The
contributions are always in odd powers of $b_\mu$. Let us discuss, for instance, the graphs linear in $b_\mu$
at an arbitrary loop order, that have, therefore, one nonminimal interaction.
They are power counting superficially cubic divergent. Consider that the momentum of the
internal photon that couples to the background vector is $k$ and thus the vertex is given by
$ie \varepsilon^{\rho \kappa \alpha \beta}b_\kappa k_\alpha \gamma_\beta$. So, the amplitude
will be given by
\be
A_{\mu \nu}= \varepsilon^{\rho \kappa \alpha \beta}b_\kappa I_{\mu \nu \rho \alpha \beta},
\ee
with $I_{\mu \nu \rho \alpha \beta}$ being a multiloop integral.

The above integral has dimension of $m^3$ and we are interested in the limit $p^2 \to 0$. Considering
its Lorentz structure, we can have all combinations of indices in two types of terms:  the first has the form
$p_\mu p_\nu p_\rho g_{\alpha \beta}$, which will give null contributions due to the antisymmetric tensor,
and the second has the form $m^2 p_\beta g_{\mu \alpha}g_{\nu \rho}$. The non-null terms are all of the
second type and will give a contribution proportional to
\be
m^2 \varepsilon_{\mu \nu \kappa \beta} b^\kappa p^\beta.
\ee

It is interesting to note that the CS-like contribution here comes from a product between the
Levi-C\`{\i}vita tensor and a term quadratic in the metric tensor, which is different from the case,
in the same loop order, where the nonminimal vertex is external. This is an effect of the extra
power of internal momentum in the numerator of the integrand. Although the factor $m^2$ suggests that
this coefficient is also originated from quadratic divergent integrals,  it is not possible, without an
explicit calculation, to assure that it is related to some gauge symmetry breaking term in the pure QED sector.
Since this CPT odd term respects gauge invariance of the action, there is
no constraint to prevent its generation in higher loop order (at first order in $b_\mu$).
The analysis above is easily extended for higher powers in the background vector $b_\mu$. For the
cubic contribution in $b$ we will have something proportional to
\be
m^4 b^2 \varepsilon_{\mu \nu \kappa \beta} b^\kappa p^\beta
\ee
and so on.

Another point that should be discussed is the fact that increasing the order in
the coupling constant will allow for higher power contributions in the background vector together with
the increasing of the degree of divergence of the integrals. Nevertheless, it was shown in \cite{Belich2}
that the magnitude of the Lorentz parameter in this model is extremely small. One condition which
is reasonable to be imposed for the recovering of QED is $|b^2| \Lambda^2 << 1$. The effect of
the divergences can be seen in a simplified form by substituting $m^2$ by $\Lambda^2$ in the
coefficients. Let us, for instance, discuss the CPT odd term. In the contribution linear in $b$
we have $\Lambda^2 \varepsilon_{\mu \nu \kappa \beta} b^\kappa p^\beta$.
For the cubic contribution in $b$, we will have
$\Lambda^4 b^2 \varepsilon_{\mu \nu \kappa \beta} b^\kappa p^\beta$,
and so on. Although the generation  of Lorentz violating terms is unavoidable beyond one-loop order,
the predictability of the effective model is assured by the cutoff inequality
above. That is because at the same loop order additional nonminimal vertices contribute with
positive powers of $b_\mu$, whereas higher loop orders
for a fixed number of nonminimal vertices is controlled by the smallness of the structure constant.

For an effective model, the cutoff energy is a very important parameter and should be established on physical grounds.
Desired features of a Quantum Field Theory like causality and stability can be lost at very high energies \cite{causality1},
\cite{causality2}. The condition imposed by the inequality $|b^2| \Lambda^2 << 1$ is such that higher power terms in $b_\mu$ does not
proliferate at a given high loop order. In the paper \cite{Belich2}, it has been set a bound such that $e \cdot |b_\mu|<10^{-32}\,(eV)^{-1}$.
This bound with the addition of the inequality we propose above guaranties that this effective model is not considered at energy
scales beyond the Planck scale.

\section{Concluding Comments}

In this paper, we have studied a modification of the Quantum Electrodynamics, which includes,
besides the minimal coupling, a nonminimal one which violates Lorentz symmetry. It is a model close
to the one analyzed in \cite{mgomes2}, where the minimal coupling has not been considered.
Besides, the addition of
the minimal coupling interaction opens the possibility of the quantum induction of a Chern-Simons-like term.

The one-loop radiative corrections to the photon two point function, in this model, has four distinct
contributions. The one from the conventional QED, the two identical crossed terms which would have
the Carroll-Field-Jackiw form  and the last one, which would be a Lorentz violating CPT even term,
with the particular form discussed in \cite{carrol} for the interaction between the gauge and the aether fields.
We have shown at one-loop order that the gauge invariance of the model fixes the coefficients of
the CPT odd and the CPT even Lorentz violating corrections to be null if a conventional gauge invariant regularization
procedure is used. However, it is always possible to choose a
gauge noninvariant procedure and then restore the gauge symmetry by means of a non-symmetric counterterm. In this case, since
the Lorentz violating and Lorentz preserving sectors are independent of each other, the Lorentz breaking terms would survive.
This would be equivalent to use different
regularization procedures in the different sectors (Lorentz invariant and Lorentz violating). Nevertheless, we think the natural framework
is choosing a unique regularization for the calculation of the integrals of the same amplitude. In this case, the calculation with a transversality
preserving one would not generate, at one-loop order, these two Lorentz violating terms (CPT-odd and CPT-even).

The analysis performed above is very close to \cite{scarp1}, where the quantum generation of
a Chern-Simons-like term in the modified QED with an axial term in the fermion sector has been
considered in the light of gauge invariance. In that case, however, it has been shown that gauge
invariance plays no role in the determination of the coefficient under study. Although being
a surface term, the $\alpha_2$ defined above, it can be left undetermined if it is used the
condition $\alpha_3=2\alpha_2$.

Whilst the use of a gauge invariant procedure yields a null coefficient for Lorentz violating terms in the pure
photon sector at one-loop order, the nonrenormalizability of the model prevent us from drawing
general statements at higher orders. However, we are still able to display the general form of
the breaking terms (in the limit $p^2 \to 0$) beyond one-loop order.

\subsection*{Acknowledgements}

This work was partially supported by CNPq and FAPEMIG. The authors wish to thank Prof. M. O. C. Gomes
and J. A. Helayel-Neto for illuminating discussions.


\end{document}